\documentstyle[preprint,aps]{revtex}
\input epsf.tex
\def\DESepsf(#1 width #2){\epsfxsize=#2 \epsfbox{#1}}
\begin{document}
\preprint{\vbox{\hbox{BNL-, UM-P-96/68}}}
\draft

\title{ Color-Octet Contribution and Direct CP Violation in 
$B\to \psi (\psi') X$ }
\author{Xiao-Gang He$^1$ and A. Soni$^2$}
\address{$^1$ School of Physics, University of Melbourne\\
Parkville, Vic 3052, Australia\\
and\\
$^2$Theory Group, Brookhaven National Laboratory, Upton, NY\ \ 11973}
\date{August, 1996}
\maketitle
\begin{abstract}
We study $c \bar c$ color-octet contribution to $B\to \psi
(\psi') X$. When this contribution is included, 
the theoretical predictions for the branching ratios 
become in much better agreement with the experiment.
This mechanism also enhances the partial rate asymmetries by 
about a factor of five. The inclusive $\psi(\psi')$ resulting from $b\to
d+{}$gluon can have asymmetry around a few percent whereas those from
$b\to s+{}$gluon has it around $4\times 10^{-4}$. The asymmetry in the
former modes should be observable, to a significance of $3\sigma$, with
about $(1-10)\times 10^8B$ mesons.

\end{abstract}
Recent experimental data from the Tevatron indicate that 
if only $c \bar c$ color-singlet
contribution is included the $\psi'$ production rate at large transverse
momentum predicted by QCD is about a factor of 30 below the experimental data.
It has been shown by Braaten and Fleming \cite{bc2} 
and Cho and Leibovich\cite{cho} 
that if 
the
$c \bar c$ color-octet
also contributes to the $\psi'$ production, 
the experimental data can be explained. 
Color-octet also has significant contribution 
to the $\psi$ production at the Tevatron\cite{cho}.
In this paper, we show that, if  this mechanism is
indeed the correct one, it also has important implications for $\psi (\psi')$ decays
of $B$ mesons, especially for CP violating particle-antiparticle partial rate
asymmetry.  The point is that the penguin graph leads to an appreciable
branching ratio ($\simeq 10^{-2}$) for $b\to s+{}$gluon. Now, in a
purely perturbative approach, the formation of $\psi (\psi')$ from the
gluon is severely suppressed. On the other hand, if Braaten {\it et
al}.'s mechanism tends to enhance the rate for the color-octet ($c\bar
c$ or gluon) to form the $\psi(\psi')$ then it can have important
consequences for direct CP violation in inclusive $B$ decays, via
$B\to\psi(\psi') + X$. 
This mechanism also enhances the branching ratios in these
decays so that they are much closer to the experimentally measured ones.

In the SM the amplitudes for $B$ decays are generated by the following effective
Hamiltonian:

\begin{eqnarray}
H_{eff}^q &=& {G_F\over \sqrt{2}}[V_{fb}V^*_{fq}(c_1O_{1f}^q + c_2 O_{2f}^q) -
\sum_{i=3}^{10}(V_{ub}V^*_{uq} c_i^u
+V_{cb}V^*_{cq} c_i^c +V_{tb}V^*_{tq} c_i^t) O_i^q] +H.C.\;,
\end{eqnarray}

\noindent where the
superscripts $u,\;c,\;t$ indicate the internal quarks, $f$ can be $u$ or 
$c$ quark. $q$ can be $d$ or $s$ quark depending on if the decay is a $\Delta S = 0$
or $\Delta S = -1$ process.
The operators $O_i^q$ are defined as

\begin{eqnarray}
O_{f1}^q &=& \bar q_\alpha \gamma_\mu Lf_\beta\bar
f_\beta\gamma^\mu Lb_\alpha\;,\;\;\;\;\;\;O_{2f}^q =\bar q
\gamma_\mu L f\bar
f\gamma^\mu L b\;,\nonumber\\
O_{3,5}^q &=&\bar q \gamma_\mu L b
\bar q' \gamma_\mu L(R) q'\;,\;\;\;\;\;\;\;O_{4,6}^q = \bar q_\alpha
\gamma_\mu Lb_\beta
\bar q'_\beta \gamma_\mu L(R) q'_\alpha\;,\\
O_{7,9}^q &=& {3\over 2} e_{q'}\bar q \gamma_\mu L b  \bar q'
\gamma^\mu R(L)q'\;,\;O_{8,10}^q = {3\over 2}e_{q'}\bar q_\alpha
\gamma_\mu L b_\beta
\bar q'_\beta \gamma_\mu R(L) q'_\alpha\;,\nonumber
\end{eqnarray}

\noindent where $R(L) = 1 +(-)\gamma_5$, 
and $q'$ is summed over $u$, $d$, $s$, and $c$.  $O_{1,2}$ are the tree
level and QCD corrected operators. $O_{3-6}$ are the strong gluon induced
penguin operators, and operators $O_{7-10}$ are due to $\gamma$ and Z exchange,
and ``box'' diagrams at loop level. The WC's $c_i^f$ are defined
at the scale of $\mu \approx
m_b$. Although  the WC's have been evaluated to the next-to-leading order in QCD \cite{nlo,dh1}, 
we will use the leading 
order WC's to be consistent with the matrix elements
which were evaluated using NRQCD to the leading order. 
We give the coefficients below for $m_t = 176$ GeV, $\Lambda_4 = 0.2$ GeV,
and $\mu = m_b = 5$ GeV,

\begin{eqnarray}
c_1 &=& -0.249\;,\;\; c_2 = 1.108\;,\;\;
c^t_3 =0.0116\;,\;\; c^t_4 =-0.0249\;,\;\;
c^t_5 =0.0073\;,
 c^t_6 =-0.0300\;,\nonumber\\
c^t_7 &=&0.0011\;,\;\; c_8^t = 0.0004\;,\;\;
c_9^t =-0.0092\;,\;\; c_{10}^t =0.0021\;, \nonumber\\
c_{3,5}^{u,c} &=& -c_{4,6}^{u,c}/N = P^c_s/N\;,\;\;
c_{7,9}^{u,c} = P^{u,c}_e\;,\;\; c_{8,10}^{u,c} = 0\;;
\label{zero}
\end{eqnarray}

\noindent where $N$ is the number of colors.
The leading contributions to $P^i_{s,e}$ are given by:
 $P^i_s = (\alpha_s/8\pi) c_2 (10/9 +G(m_i,\mu,q^2))$ and
$P^i_e = (\alpha_{em}/9\pi)(N c_1+ c_2) (10/9 + G(m_i,\mu,q^2))$.  
The function $G(m,\mu,q^2)$ is given by

\begin{eqnarray}
G(m,\mu,q^2) = 4\int^1_0 x(1-x) \mbox{d}x \mbox{ln}{m^2-x(1-x)q^2\over
\mu^2}\;,
\end{eqnarray}

\noindent where $q$ is the momentum carried by the virtual gluon in 
the penguin
diagrams. When $q^2 > 4m^2$, $G(m,\mu,q^2)$ becomes  complex. 
In our calculation, we will
use $m_u = 5$ MeV and $m_c = 1.35$ GeV\null. Although there is
considerable uncertainty in the effective values of these masses,
especially so for $m_u$, our results are rather insensitive to the
numerical values of these masses.

Using factorization approximation, we have

\begin{eqnarray}
M(B\to J/\psi X_s)&=& {G_F\over \sqrt{2}}[A_1<X^1_s|\bar s 
\gamma_\mu (1-\gamma_5) b|B>
<\psi (\psi')|\bar c \gamma^\mu (1-\gamma_5) c|X^1>\nonumber\\
&+& 2A_8<X^8_s|\bar s \gamma_\mu (1-\gamma_5)T^a b|B>
<\psi (\psi')|\bar c \gamma^\mu (1-\gamma_5)T^a c|X^8>\nonumber\\
A_1&=&[V_{cb}V_{cs}^*(c_1 +{c_2\over N} - c_3^{cu} - {c_4^{cu}\over N} - 
c_5^{cu} - {c_6^{cu}\over N} - c_7^{cu} - {c_8^{cu}\over N} - c_9^{cu} 
- {c_{10}^{cu}\over N})\nonumber\\
&-&V_{tb}V_{ts}^*( c_3^{tu} + {c_4^{tu}\over N} + c_5^{tu} + {c_6^{tu}\over N} 
+ c_7^{tu} + 
{c_8^{tu}\over N} + c_9^{tu} + {c_{10}^{tu}\over N})]\nonumber\\
A_8&=&V_{cb}V_{cs}^*(c_2 - c_4^{cu} - c_{6}^{cu} - c_{8}^{cu}  - c_{10}^{cu})
-V_{tb}V_{ts}^*( c_4^{tu} +  c_6^{tu} + c_8^{tu} + c_{10}^{tu})\;.
\label{amp}
\end{eqnarray}

\noindent where $c_i^{cu} = c_i^c - c_i^u$ and  $c_i^{tu} = c_i^t - c_i^u$,
$X_s^1+X^1 = X_s^8+X^8 = X_s$.  The term proportional to
$A_1$ is the color-singlet amplitude and the  term
proportional to $A_8$ is the color-octet amplitude.

If the color-octet contribution is neglected, the branching ratios for 
$B\rightarrow \psi (\psi') X$
are too small compared with the experimental values. In order to reproduce the 
experimental data, the number of
colors $N$ is traditionally treated as a free parameter to parametrize the 
non-factorizable and the color-octet contributions.
The effective number of colors $N$ is then determined from
$B\rightarrow \psi (\psi') X$ to be close to 2\cite{n2}. 
This does not really identify 
where the new contributions come from. However, if the color-octet effects
identified above have significant contributions, one
may not need to treat
$N$ as a free parameter; $N=3$ as given by QCD 
may work fine. 
The color octet mechanism with $N=3$ indeed improves the
situation significantly.
This has been 
pointed out by Ko, Lee and Song\cite{lk}. 
Our calculations confirm their results.
In their work the penguin contributions
are not included. When penguin contributions are included, they have important
implications for direct CP violation in particle-antiparticle partial rate
asymmetries in these decays
although their effect on the absolute rates is minimal. 

In order to obtain non-zero partial rate asymmetry,
it is necessary to  have non-zero CP violating phases and CP conserving strong
phases due to the final state  rescattering for different amplitudes. In the
above case, the CP violating phases are provided by the phases in the CKM
matrix elements $V_{cb}V_{cs}^*$ and $V_{tb}V_{ts}^*$. We will use the Wolfenstein
parametrization such that the element $V_{ub}$ is given by $|V_{ub}|e^{-i\gamma}$. 
For  the strong phases,
we will use the phases generated at the quark level by appealing to quark-hadron
duality. This should be at least a good indication for the size of the phases \cite{soni}.
Naively the strong phases are generated by exchanging $u$ and $c$ quarks
in the loop. However, CPT theorem dictates that the phases generated by 
the $c$ quark
in the loop not to contribute to the rate asymmetry 
for the production of $\psi(\psi')$\cite{cpt,more,dhp}. Both the
strong penguin WC's $c^u_{3,4,5,6}$ and the electroweak penguin WC's $c^u_{7,8,9,10}$
contribute to the strong phases.  It is, however, interesting to note that
if the color-octet amplitude is neglected, the strong penguin do not generate
non-zero strong phases because the combination $c_3^u + c^u_4/N 
+c_5^u +c_6^u/N$ in $A_1$ is identically zero as can be seen from Eq.\ref{zero}.
The leading non-zero  strong phases are then generated by electroweak penguin
and are therefore small. If these phases are the only ones, the partial rate
asymmetry is predicted to be very small as shown in  Figure~1 \cite{dhp}. Here
we have used  $N=2$ since no color-octet contributions 
have so far been included as discussed before.
If the color-octet contribution turns out to be significant, the situation can
become dramatically different. 

Including the color-octet contribution, we have

\begin{eqnarray}
|M(B\to \psi (\psi') X_s)|^2&=& G_F^2 Tr(\not P_s + m_s)\gamma^\mu (\not
P_b + m_b) \gamma^\nu(1-\gamma_5)\nonumber\\
&\times&(-g_{\mu\nu} + {P^{\psi(\psi')}_\mu P^{\psi(\psi')}_\nu\over 
m^2_{\psi(
\psi')}})
{2m_c<O^{\psi(\psi')}_1(^3S_1)>\over 3}\nonumber\\
&\times&[|A_1|^2
+ {2\over N}|A_8|^2{<O^{\psi(\psi')}_8(^3S_1)>\over <O^{\psi(\psi')}_1(^3S_1)>}]\;,
\label{m}
\end{eqnarray}
where the operators $O^{\psi(\psi')}_{1,8}(^3S_1)$ are defined
in Ref.~\cite{bc1,bbl}.

From eq.\ref{m}, the branching ratios and the CP violating
partial rate asymmetries can be easily calculated. 
To quantitatively assess the importance of  the color-octet contribution  
we recall that it  has been shown by  Ref.~\cite{bc2,cho} that even with a small
matrix element for color-octet to produce a $\psi(\psi')$, the experimental data from 
the Tevatron can be understood. Fitting the experimental data from the Tevatron,
 Cho and Leibovich obtain\cite{cho} 

\begin{eqnarray}
<O^{\psi}_8(^3S_1)> &=& 1.2\times 10^{-2} \mbox{GeV$^3$}\;,\nonumber\\
<O^{\psi'}_8(^3S_1)>&=& 7.3\times 10^{-3} \mbox{GeV$^3$}\;.
\end{eqnarray}
The color-singlet matrix elements determined from leptonic decays of $\psi$ and $\psi'$
are\cite{lk}
 
\begin{eqnarray}
<O^{\psi}_1(^3S_1)> = 1.32 \mbox{GeV$^3$}\;,\nonumber\\
<O^{\psi'}_1(^3S_1)> = 0.53 \mbox{GeV$^3$}\;.
\end{eqnarray}

Without the color-octet contributions, 
and with $N=3$, the
branching ratios predicted, are several
times smaller than the experimental values: 
 $Br(B\rightarrow \psi X) = (0.8\pm 0.08)\%$, and
$Br(B\rightarrow \psi' X) = (0.34\pm 0.05)\%$. 
The inclusion of the color-octet contributions improves the situation with the 
branching ratios now predicted to be:
 $Br(B\rightarrow \psi X) = 0.54\%$, and
$Br(B\rightarrow \psi' X) = 0.25\%$, with $N = 3$.
These numbers are in good agreement with Ko et al \cite{lk} and they are also 
much closer to the experimental
values. 

Unfortunately the results are very sensitive to $c_1$ and $c_2$. The dominant color-singlet contributions 
are from operators $O_{1,2}$ which are proportional to $c_1+c_2/N$. There is an accidental 
cancellation here. Had one used the next-to-leading coefficient for $c_{1,2}$, the cancellation
is even more severe. If one adjusts the scale $\mu$ and $\Lambda_4$ 
for the leading coefficient, one can get larger
values for the branching ratios. There are other uncertainties 
in the evaluation of the branching ratios,
namely, there are more operators which may 
contribute to the branching ratio. For example,
there may be contributions from $O_{1,8}^{\psi(\psi')}(^1S_0)$. The value for 
$<O_1(^1S_0)>$ is expected to be
much smaller than $<O_1(^3S_0)>$. Its contribution to the branching ratios 
is expected to be small. The
contributions from $O_8(^1S_0)$ may be potentially large because $<O_8(^1S_0)>$ may be not too much
smaller than $<O_8(^3S_0)>$. If this is indeed the case, 
the experimental branching ratios can be
easily reproduced. 

When the color-octet contributions are included the strong penguin also
generate strong phases through the coefficient $A_8$ in Eq.~\ref{amp}. These
phases are much larger than the ones generated by the electroweak penguins,
and therefore much larger partial rate asymmetries result. The 
results are shown in Figure~2. In the figure, we used $q^2 = m_{\psi(\psi')}^2$
because the $\psi(\psi')$ carries most of the momentum from the virtual gluon.
We also set $\alpha_s$ at  $q^2=m^2_{\psi(\psi')}$ and the corresponding value
$\alpha_s(m^2_{\psi(\psi')})= 0.27$ for $\Lambda_4 = 0.2$ GeV. If larger  $\alpha_s$ is used, the
asymmetries become bigger. 
The asymmetry
for $B\to \psi' X_s$ is slightly larger than that for  $B\to \psi X_s$. 
This is due to the fact that the ratio of the color-octet matrix
element to color-singlet is larger for the $\psi'$. 
We also considered
the contribution from dipole penguin operators, $O_{11} = (g_s/16 \pi^2)m_b
\bar s \sigma_{\mu\nu}RT^ab G^{\mu\nu}_a$ and $O_{12} = (e/16 \pi^2)m_b \bar
s \sigma_{\mu\nu}Rb F^{\mu\nu}_a$. Here $G^{\mu\nu}$ and 
$F^{\mu\nu}$ are the gluon and photon field strengths, respectively. It has
been shown that the operator $O_{11}$ can enhance certain $B$ decay branching
ratios  by as much as 30\% \cite{dht}. However, its
contribution to $B\to \psi(\psi') X_s$ branching ratio is less than $10^{-4}$
and to partial rate asymmetry is less than $10^{-5}$. $O_{12}$ contributions
are even smaller. We remark that even if the operator $O_8(^1S_0)$  
contributes significantly to the branching ratios, it will not 
introduce new strong phases 
in the 
amplitude because its contributions are 
proportional to $(c_4^{u(t)c}-c_6^{u(t)c})$ which generate
vanishing absorptive amplitudes. And therefore, it will not affect the asymmetries evaluated here.

It is clear from 
comparison of Fig.1 and Fig.2 that inclusion of the color-octet enhances the asymmetries. 
The asymmetries are bigger by about a factor of five.
At present the CP violating phase $\gamma$ is not well determined; $\sin\gamma$
can vary from 0.1 to 1. If we use the best fit value from the experimental data,
$\gamma$ is about $70^\circ$ \cite{ali}. 
With this value, the rate asymmetries for $B\rightarrow \psi (\psi')$
are about $4\times 10^{-4}$ ($6\times 10^{-4}$). 
In order to observe the asymmetries at the 3$\sigma$ level 
in $B\rightarrow \psi X_s$ and 
$B\rightarrow \psi' X_s$, one would need about 
$4\times 10^{9}$ $B$ decays. 
This number does not include any factor(s) for 
experimental efficiencies. So the number of B's needed
is likely to be even higher depending on the specific final states of
the $\psi(\psi')$ that are accessible.  

The situation with $B\to \psi(\psi') X_d$ 
($X_d$ denotes states without strangeness
number) is better. The analysis is similar to $B\to \psi(\psi') X_s$ case.
One only  needs to change the relevant CKM matrix elements $V_{cb}V_{cs}^*$
and $V_{tb}V_{ts}^*$ to $V_{cb}V_{cd}^*$ and $V_{tb}V_{td}^*$, respectively
in Eq.~\ref{amp}. The results are shown in  Figure~3. The asymmetries can be
as large as 1.5\%. They are about 1\% (1.5\%) for 
$B\rightarrow \psi (\psi') X_s$ with $\gamma = 70^\circ$. 
It is interesting to observe that these asymmetries 
are similar to those obtained in Ref.\cite{soares} by considering absorptive 
contribution from rescattering of $c \bar c$ color-octet states.
For the $X_d$ final state the branching ratios are smaller by a factor of
$|V_{cd}/V_{cs}|^2$ compared with $B \to \psi(\psi') X_s$. Therefore to observe 
the asymmetries in $B\to \psi(\psi') X_d$ at the
$3\sigma$  level, one would need about $1\times 10^{8}$ $B$ decays.
Assuming an effective efficiency of 0.2 (i.e. including
the branching ratio into some specific final state(s))
would make the actual number be
more like $5\times 10^{8}$.

The number of B's needed is clearly rather large so that even 
$B$ factories may have a difficult time. On the other hand the $\psi$
and $\psi'$ tend to give distinctive signal which may even be accessible
in a hadronic environment with a B-detector. 

In the above, we have considered the parton level processes, $b\rightarrow s(d) \psi (\psi')$.
If one considers the parton level processes, $b\rightarrow s(d) Y$ with $Y$ being 
$\psi$ and other $c\bar c$ states which can 
materialize into $\psi$, and do not isolate each
individual $Y$ and let it decay into $\psi$, then the hadron level process, $B\rightarrow 
\psi X$ will have more sources to provide strong phases. In addition to the parton level
strong phases discussed above, there is also the possiblilty
of generating strong phases 
from resonant effects\cite{resonances}. The partial rate asymmetries may be 
even larger than what we obtained here. We will discuss 
the results from these mechanisms elsewhere.

One might think that the same mechanism will also enhance the rate asymmetry
in $B\to \eta_c X_s (X_d)$. It turns out that this is not true 
here if only $O_{1,8}(^3S_0)$
operators are included. In this
case, for the same  reason as for $B\to \psi (\psi') X$, the color-singlet
only generates strong phases  through electroweak penguins. However, the
inclusion of color-octet contributions will not improve the situation because
the analogous color-octet parameter $A_8(\eta_c)$ is different than the $A_8$
parameter in Eq.\ref{amp}. The strong penguin contribution in $A_8(\eta_c)$
is proportional
to $(c_4^{u(t)c}-c_6^{u(t)c})$ which 
generates vanishing strong phase too. However, if
the operator $O_8(^1S_0)$ contributes significantly, strong phases will
be generated by the strong penguins because the contributions are 
then proportional
to $(c_4^{u(t)c}+c_6^{u(t)c})$.  There may be sizable rate asymmetries
which also need further study.
\bigskip\bigskip

XGH would like to thank Dr. Ma for useful discussions. 
XGH was supported by Australian Research Council, and 
AS was supported in part by the U.S. DOE contract
DE-AC-76CH00016. XGH would like to thank the Theory Group at the Brookhaven National 
Laboratory for hospitality where this work was started.

\begin{figure}[htb]
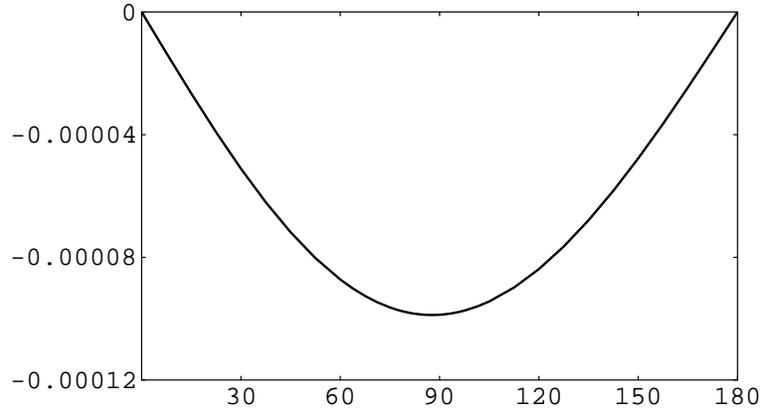

\centerline{ \DESepsf(soni1.epsf width 10 cm) }
%\vglue 4in
\smallskip
\caption {The partial rate asymmetry for $ B\to \psi (\psi') X_s$ without color-octet
contribution with $|V_{cb}| = 0.04$, $|V_{ub}/V_{cb}|=0.08$ and $|V_{us}| = 0.22$. 
The vertical axis is the asymmetry and the horizontal axis 
is the value in degree for the phase angle
$\gamma$ in the Wolfenstein parametrization.}
\label{gamma}
\end{figure}

\begin{figure}[htb]
\centerline{ \DESepsf(soni2.epsf width 10 cm) }
%\vglue 4in
\smallskip
\caption {The partial rate asymmetry for $ B \to \psi (\psi') X_s$. The solid and dashed
lines are for $B\rightarrow \psi X_s$, and $B\rightarrow \psi' X_s$, respectively.
}
\label{gamma1}
\end{figure}

\begin{figure}[htb]
\centerline{ \DESepsf(soni3.epsf width 10 cm) }
%\vglue 4in
\smallskip
\caption {The partial rate asymmetry for $B\rightarrow \psi (\psi') X_s$.
The solid and dashed
lines are for $B\rightarrow \psi X_d$, and $B\rightarrow \psi' X_d$, respectively}
\label{gamma2}
\end{figure}

\end{document}